\documentclass[twocolumn]{emulateapj}

\usepackage{latexsym,amssymb}
\usepackage{amsmath}
\usepackage{natbib}

\def\lsim{\mathrel{\rlap{\lower4pt\hbox{\hskip1pt$\sim$}}
    \raise1pt\hbox{$<$}}}                
\def\gsim{\mathrel{\rlap{\lower4pt\hbox{\hskip1pt$\sim$}}
    \raise1pt\hbox{$>$}}}                

\usepackage[colorlinks=true,linkcolor=blue,citecolor=blue]{hyperref}

\slugcomment{Draft \today}

\shorttitle{Necessary Conditions for Short Gamma-Ray Burst Production}

\shortauthors{Murguia-Berthier et al.}

\begin{document}

\title{Necessary Conditions for Short Gamma-Ray Burst Production in Binary Neutron Star Mergers}

\author{Ariadna Murguia-Berthier\altaffilmark{1,2}, Gabriela Montes\altaffilmark{1}, Enrico Ramirez-Ruiz\altaffilmark{1}, Fabio De Colle\altaffilmark{3} and William H. Lee\altaffilmark{4}}
\altaffiltext{1}{Department of Astronomy and
  Astrophysics, University of California, Santa Cruz, CA
  95064}
\email{gmontes@ucsc.edu}
  \altaffiltext{2}{Facultad de Ciencias, Universidad Nacional Aut\'onoma
de M\'exico, Distrito Federal,
M\'exico 04510}
\altaffiltext{3}{Instituto de Ciencias Nucleares, Universidad Nacional Aut{\'o}noma de M{\'e}xico, A. P. 70-543 04510 D. F. Mexico}
\altaffiltext{4}{Instituto de Astronom\'ia, Universidad Nacional Aut{\'o}noma de M{\'e}xico, A. P. 70-264 04510 D. F. Mexico}

\begin{abstract}
The central engine of short gamma-ray bursts (sGRBs) is hidden from direct view, operating at a scale much smaller than that probed by the emitted radiation. Thus we must infer its origin not only with respect to the formation of the {\it trigger} - the actual astrophysical configuration  that is capable of powering a sGRB - but also from the consequences that follow from the various evolutionary pathways  that may be involved in producing it. Considering binary neutron star mergers we critically evaluate, analytically and through numerical simulations, whether the neutrino-driven wind produced by the newly formed hyper-massive neutron star can allow the collimated relativistic outflow that follows its collapse to actually produce a sGRB or not. Upon comparison with the observed sGRB duration distribution, we find that collapse cannot be significantly delayed ($\leq 100$~ms) before the outflow is choked, thus limiting the possibility that long-lived hyper-massive remnants  can account for these events. In the case of successful breakthrough of the jet through the neutrino-driven wind, the energy stored in the cocoon could contribute to the precursor and extended emission observed in sGRBs.
\end{abstract}

\section{Introduction}
The most popular  model for short gamma-ray bursts (sGRBs) invokes the coalescence of binary neutron stars
and the subsequent  production of  a beamed, relativistic outflow \citep{eichler89, paczynski91, narayan92, meszaros92}.
The launching of a relativistic  jet requires  material with sufficient free energy to escape the gravitational field of the central object as well as a mechanism for imparting some directionality to the outflow \citep{mochkovitch93, rosswog03, aloy05, rezzolla11,palenzuela2013}.  
A potential death trap for such relativistic outflows is the amount of entrained baryonic mass from the surrounding environment \citep[see e.g.][and references therein]{lee07}.  
In neutron star binaries the elevated post-merger  neutrino fluxes are capable of ablating  matter from the surface of the remnant  at a  rate
\begin{equation}
\dot{M}_{\rm w}\approx 5 \times 10^{-4}\left({L_\nu \over 10^{52}\;{\rm erg/s}}\right)^{5/3}\;M_\odot/{\rm s}
\label{eq:wind}
\end{equation}  
\citep{qian96, rosswog02, dessart09}.
Thus the rest mass flux arising from the neutrino-driven wind bounds the bulk Lorentz factor of the jet to
\begin{equation}
\Gamma_\nu \approx 10 \left({L_{\rm jet}\over 10^{52}\;{\rm erg/s}}\right)\left({\dot{M}_{\rm w} \over   5 \times 10^{-4}\;M_\odot/{\rm s}}\right)^{-1}
\label{eq:bpop}
\end{equation}  
and hence the successful launch of a highly relativistic jet might have to wait until the collapse  of the merger remnant  and the ensuing formation of the black hole plus debris disk system. 

The fate  of the post-merger, hyper-massive neutron star  is, however, uncertain, and is contingent  on the mass limit for support of a hot, differentially rotating configuration
\citep[e.g.][]{baumgarte2000,duez2006,giacomazzo13,hotokezaka13a}.
The threshold for collapse can be calculated roughly as $M_{\rm thres}=1.35 M_{\rm cold}$
\citep{shibata06}, where $M_{\rm cold}$ is the corresponding value for a  cold,  non-rotating configuration. In agreement  with the mass determination in PSR J0348+0432
\citep[$M_{\rm cold} \gtrsim 2 M_\odot$;][]{demorest10}, 
a total mass greater than $\approx 2.7 M_\odot$ is required for prompt collapse to a black hole. When $M_{\rm cold} < M < M_{\rm thres}$, various mechanisms  could act to dissipate and/or transport energy and angular momentum, possibly inducing collapse after a delay which could range from tens of milliseconds  to a few seconds \citep[for a recent review see][]{faber12}.
During this period, a baryon loaded wind is continuously ejected for a time $t_{\rm w}$, which precedes  jet formation \citep{lehner12}.
 As a result, a dense wind remains to hamper the advance of the jet, whose injection lifetime, $t_{\rm j}$,  is determined by the  viscous  time scale of the neutrino-cooled disk \citep{lee2004,lee2005a,Setiawan2004,metzger2008,lee2009}. 
  
 In this {\it Letter}, we study with the use of hydrodynamical simulations how the expansion of the relativistic  jet is modified by the previously ejected  wind and 
investigate the conditions necessary for successful sGRB production in binary neutron star mergers.  There are three sections. Section~\ref{ana}  gives an account of the properties that determine the  advancement  of the jet in the  wind and its possible successful  emergence. 
Section~\ref{num} describes the results of the numerical calculations.  Finally, Section~\ref{dis} gives a compendium of the types of observational signatures  expected for jets propagating  through neutrino-driven winds of  different mass loading and durations together with a model  proposal for generating sGRBs with precursor and  extended emission.

\section{Understanding Jet Propagation and Confinement}\label{ana}    
The properties of the neutrino-driven wind have an important effect on  a jet propagating through it.   If the isotropic equivalent  power in the jet, $L_{\rm j}$, is roughly conserved and stationary, then we balance momentum fluxes  at the working surface to obtain  \citep{matzner03,bromberg11}
\begin{equation}
\beta_{\rm h}=\frac{\beta_{\rm j}}{1+\tilde{L}^{-1/2}},
\end{equation}
where $\beta_{\rm h}$ and $\beta_{\rm j}$ are the head and shock velocities respectively, and both the shocked jet material and the shocked wind material  advance with a jet head Lorentz factor $\Gamma_{\rm h} \ll \Gamma_{\rm j}$ \citep{ramirez-ruiz2002}. Here
\begin{equation}
\tilde{L}=\frac{\rho_{\rm j}}{\rho_{\rm w}}h_{\rm j}\Gamma_{\rm j}^2,
\end{equation}
where $\rho_{\rm w}$ denotes the density of the wind material and $h_{\rm j}$ the specific enthalpy of the jet.

 \begin{figure}[t]
\centering
\includegraphics[height=0.7\linewidth,clip=true]{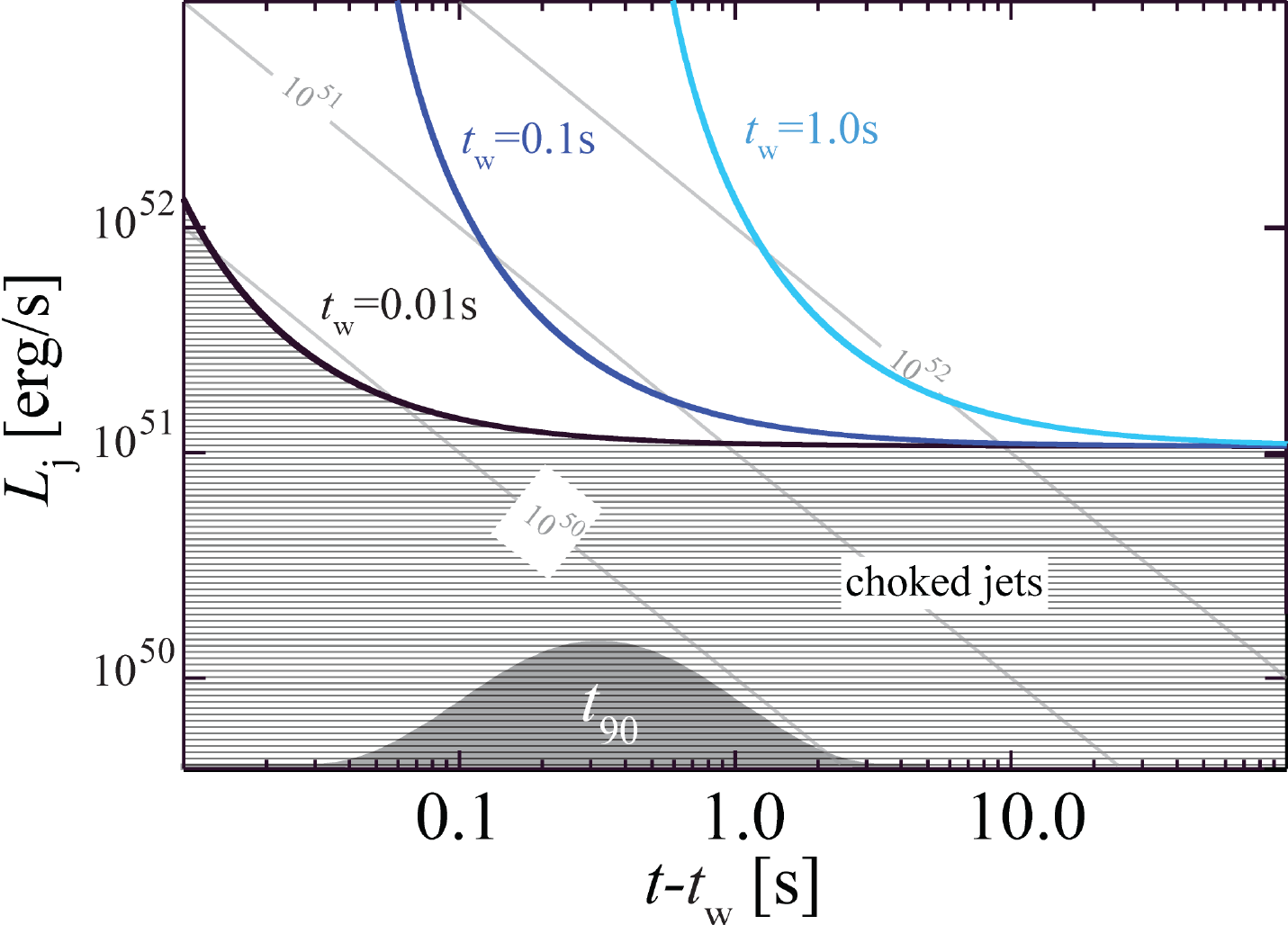}
\caption{This plot illustrates  how the jet expansion would be affected by the properties of the wind which it propagates. The axes (logarithmic) are $L_{\rm j}$ versus $t-t_{\rm w}$, where 
$t-t_{\rm w}=0$  denotes the time when the wind injection terminates and the jet starts its expansion. The  lines represent the time it takes for the jet to  break free ($t_{\rm j,b}$) from a wind with $\dot{M}_{\rm w}=10^{-3}M_\odot \, {\rm s}^{-1}$ and  $\beta_{\rm w}=0.3$ for $t_{\rm w}$ =0.01, 0.1, 1~s. Lines of constant (isotropic equivalent) jet energy  are  plotted as grey solid lines (we have assumed $h_{\rm j}=1$). When $t_{\rm w}>0.1$ s, jet injection times in excess of sGRB durations, illustrated here by the  $t_{90}$ distribution  \citep{nakar2007,gehrels2009}, are required in order to breakout  from the wind.}
\label{fig:lvst}
\end{figure}	

In a wind medium such that $\rho_{\rm w}=\dot{M}_{\rm w}/(4\pi \beta_{\rm w}  r^{2} c)$, the jet head's velocity is steady, assuming the opening angle of the jet remains constant.  Thus, the breakout time for a steady  jet  propagating through a wind with a velocity $\beta_{\rm w}$, injected during a time $t_{\rm w}$ is given by
\begin{equation}
  t_{\rm b}=t_{\rm w}\frac{\beta_{\rm h}}{\beta_{\rm h}-\beta_{\rm w}},
\end{equation}
where $t=0$ denotes the wind injection time. As the jet makes its way out and its  rate of advance is slowed down, most of the energy output during that period is deposited into a cocoon surrounding it \citep{ramirez-ruiz2002}. The jet would be  expected to break free of  the  dense wind, and in principle lead to a successful sGRB, provided the central engine feeding time $t_{\rm j}$  exceeds 

\begin{equation}
t_{\rm j, b}=t_{\rm b}-t_{\rm w}=t_{\rm w}\frac{\beta_{\rm w}}{\beta_{\rm h}-\beta_{\rm w}}.
\label{eq:tbl}
\end{equation}

Fig.~\ref{fig:lvst} shows the luminosity required  for a jet  to  breakout  from a wind medium with $\dot{M}_{\rm w}=10^{-3}M_\odot \, {\rm s}^{-1}$ and $\beta_{\rm w}=0.3$. Three cases are depicted  for $t_{\rm w}$= 0.01, 0.1, 1~s.  The most favorable region for shocks producing highly variable $\gamma$-ray light curves is above the edge of the wind, while the radiation emitted by shocks occurring below it would  be drastically absorbed as the optical depth across the wind  is enormous. If  collapse to a black hole is significantly  delayed  (i.e. $t_{\rm w}>0.1$ s), the majority of jets, with the exception of some very luminous ones, will not breakout  during the typical duration of a sGRB ($t_{90}$). If, on the other hand,  collapse occurs  more promptly (i.e. $t_{\rm w} \leq 0.1$ s),  the  successful break-through of the jet would  lead to a  detectable sGRB followed by a standard afterglow.  It is not necessary for  the head of the jet to move at  a high bulk Lorentz factor as it transverses the dense wind.  As long as the emergent outflow has a high enthalpy per baryon, it will expand and achieve its high terminal speed some distance from the edge of the wind. 

 \begin{figure}[t]
\centering
\includegraphics[height=0.75\linewidth,clip=true]{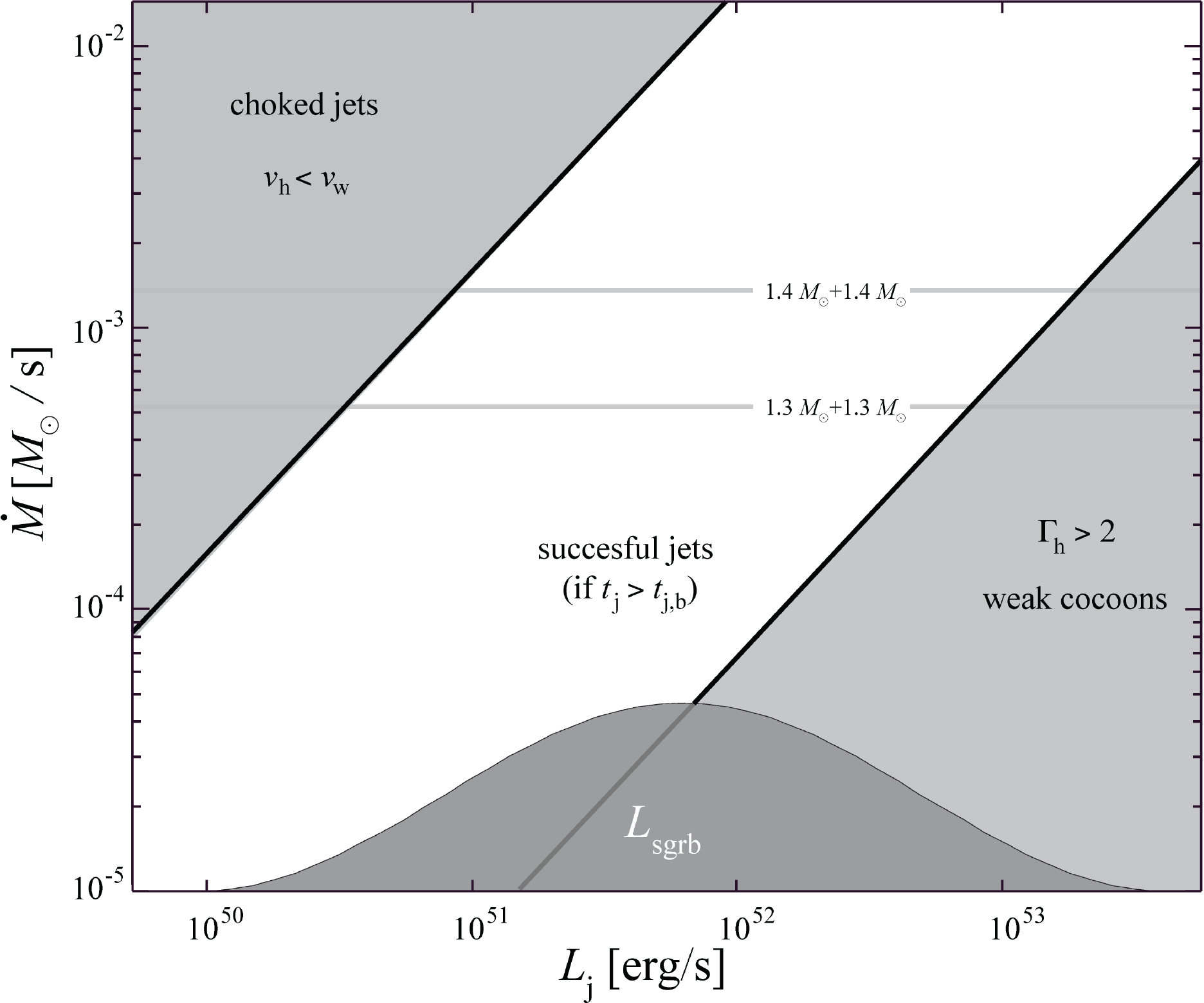}
\caption{Characteristic behavior of relativistic  jets expanding within a neutrino-driven wind.  $L_{\rm j}/\dot{M}_{\rm w}$ controls the rate of advance  of the head of the jet while $t_{\rm w}/t_{\rm j}$ determines whether or not the jet would be able to break free of the neutrino driven wind. For comparison, the  distribution of isotropic equivalent luminosities for observed sGRB is plotted \citep{berger2013,margutti2013} together with an estimate of $\dot{M}_{\rm w}$ for a 1.4 $M_\odot$ + 1.4 $M_\odot$ neutron star merger \citep{dessart09}. The mass loss rate expected for a 1.3 $M_\odot$ + 1.3 $M_\odot$ merger is calculated here using the relationship  between $L_\nu$  and post-merger remnant mass derived in \citet{rosswog03}. Similar  values for $\dot{M}_{\rm w}$ are estimated using magnetically-driven wind models \citep{siegel2014}. }
\label{fig:diag}
\end{figure}	

\begin{figure*}[]
\centering
\includegraphics[height=0.5\linewidth,clip=true]{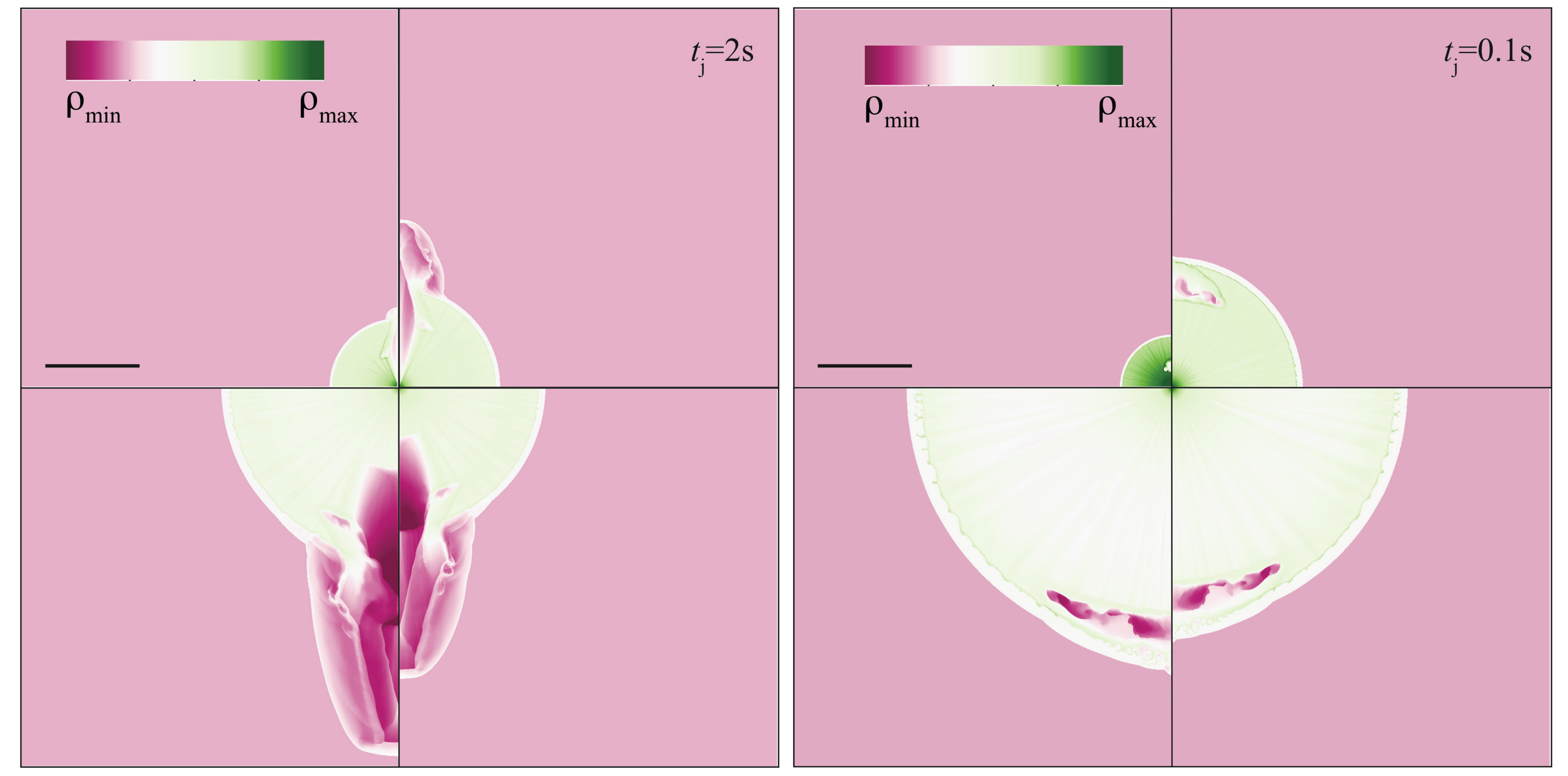}
\caption{The evolution of a  sGRB jet in  a steady neutrino-driven wind with $\dot{M}_{\rm w}=10^{-3} \rm{M_\odot \, s^{-1}}$ and $\beta_{\rm w}=1/3$. The jet is characterized by  $L_{\rm j}=10^{51}\,\rm{erg\,s^{-1}}$, $\theta_{\rm j}=10^\circ$ and $\Gamma_{\rm j}=10$. Two illustrative cases are shown for $t_{\rm j,b} < t_{\rm j}=2$~s (left panel) and  $t_{\rm j,b} > t_{\rm j}=0.1$~s (right panel).  Shown are logarithmic density contours  $[\rho_{\rm min}, \rho_{\rm max}]=[2 \times 10^{-5}, 7.0]$ in g cm$^{-3}$ together with a $3\times 10^{10}$ cm scale bar. Each snapshot has been rotated by $\pi/2$, where $t=2,~3, ~4.5,~5.5$ s (left panel) and $t=1.5,~4, 7,~8.5$~s (right panel). Calculations were done in two-dimensional cylindrical coordinates using an adaptive  grid of physical size $l_{\rm r} = l_{\rm z} = 1.2\times 10^{11}$ cm, with $300 \times 300$ cells on the coarsest grid and 7 levels of refinement, which corresponds to a maximum resolution of $6.25\times 10^6$ cm.}
\label{fig:time}
\end{figure*}
 
\begin{figure*}[]
\centering
\includegraphics[height=0.5\linewidth,clip=true]{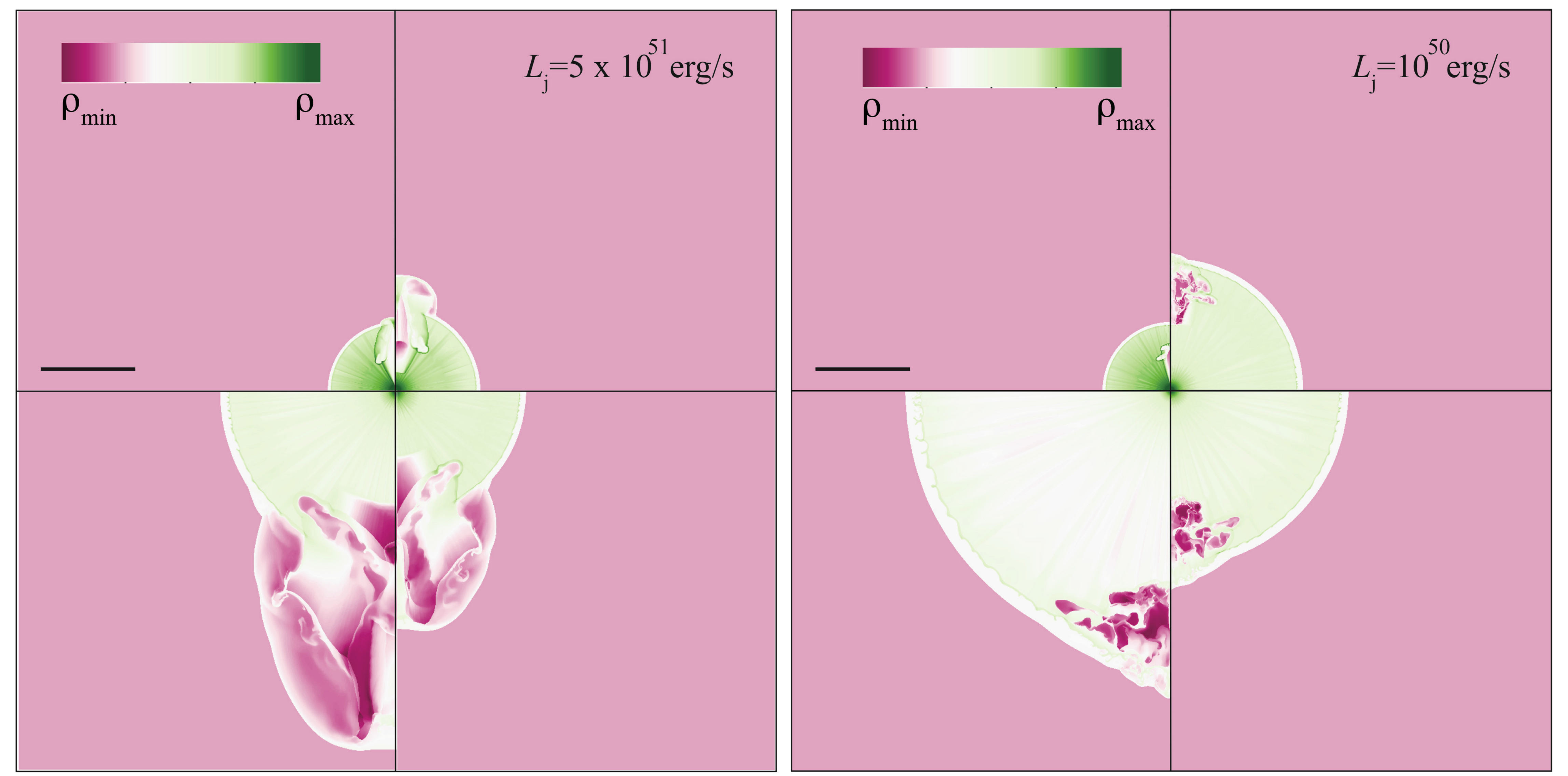}
\caption{The evolution of sGRB jets  of varying power interacting with a  wind with $\dot{M}_{\rm w}=10^{-3} \rm{M_\odot \, s^{-1}}$ and $\beta_{\rm w}=$1/3.  The numerical setup is identical to that used in Fig.~\ref{fig:time}. The jet is characterized by  $t_{\rm j}=1$s, $\theta_{\rm j}=10^{\circ}$ and $\Gamma_{\rm j}=10$. Two examples are depicted for   $L_{\rm j}=5\times 10^{51}\,\rm{erg\,s^{-1}}$ (left panel) and  $L_{\rm j}=10^{50}\,\rm{erg\,s^{-1}}$ (right panel). Each snapshot has been rotated by $\pi/2$, where $t=2,~2.5,~4.5,~5.5$ s (left panel) and $t=2, 4, 6, 8$ s (right panel).  Plotted  are the logarithmic density contours  $[\rho_{\rm min}, \rho_{\rm max}]=[2 \times 10^{-5}, 7.0]$ in g cm$^{-3}$ as well a corresponding  $3\times 10^{10}$ cm scale bar. }
\label{fig:lum}
\end{figure*}

Irrespective of $t_{\rm j}/t_{\rm w}$, a choked jet  will invariably result if $\beta_{\rm h}/\beta_{\rm w} <1$. This condition can be rewritten as
\begin{equation}
L_{\rm j} \lsim 1.1 \times 10^{51} \left({\dot{M}_{\rm w} \over    10^{-3}\;M_\odot/{\rm s}}\right) \;{\rm erg/s}. 
\label{eq:lumreq}
\end{equation}  
By contrast, relativistic expansion of the jet head within the wind medium would be guaranteed if $\tilde{L} \gg 1$,  which occurs when 
\begin{equation}
L_{\rm j} \gg 5.94 \times 10^{51}  \, \left({\dot{M}_{\rm w} \over     10^{-3}\;M_\odot/{\rm s}}\right) \;{\rm erg/s}. 
\end{equation}  
The dependence of the  jet propagation behavior   on $L_{\rm j}/\dot{M}_{\rm w}$ is summarized in Fig. \ref{fig:diag}.

\section{Numerical Simulations}\label{num}
 
\begin{figure}[]
\centering
\includegraphics[height=1.4\linewidth,clip=true]{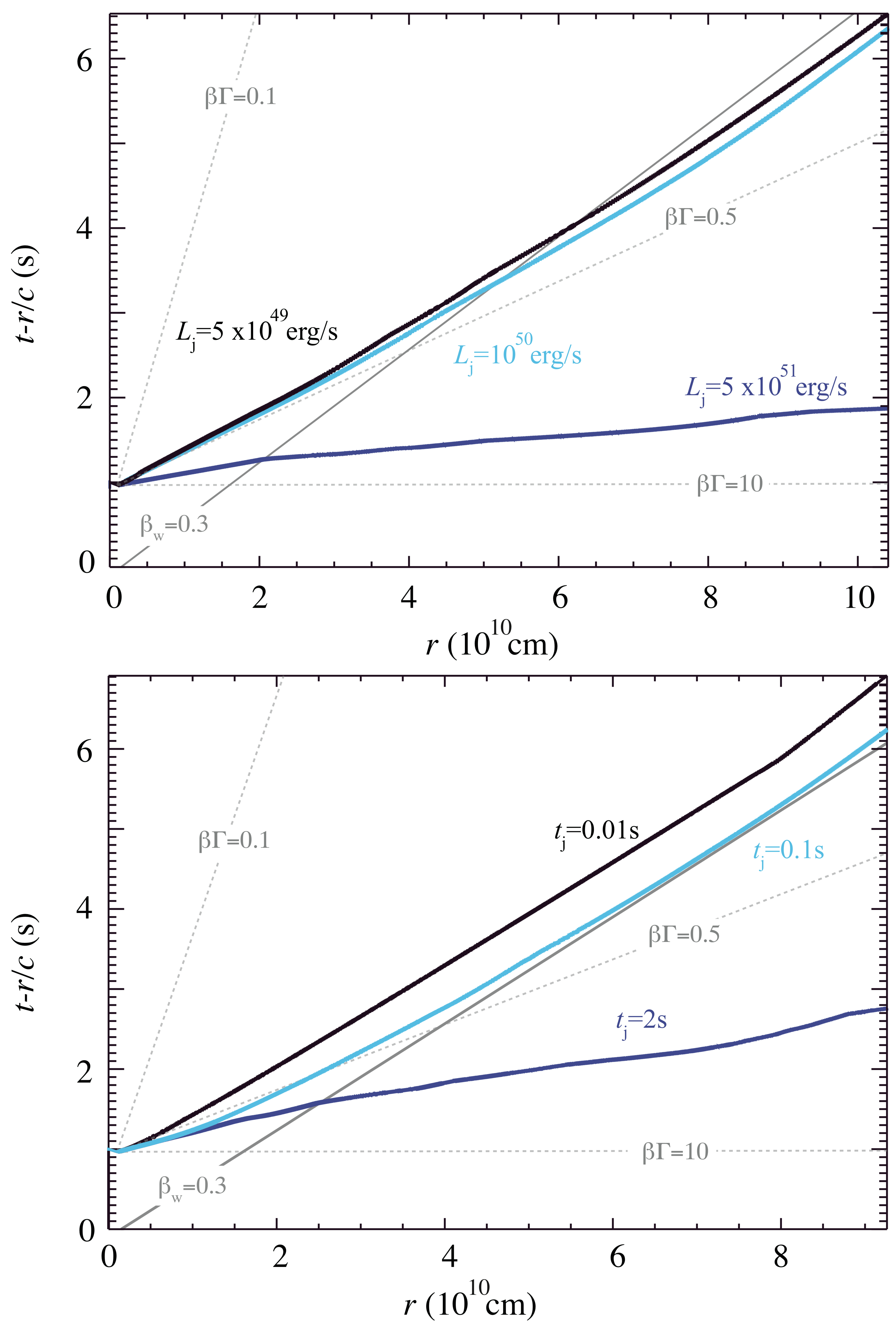}
\caption{Space time diagram in source frame coordinates of a relativistic jet propagating through a wind medium with $\dot{M}_{\rm w}=10^{-3} \rm{M_\odot \, s^{-1}}$, $t_{\rm w}=1$ s and $\beta_{\rm w}$=1/3. The axes are $r$ versus $t -r/c$, where $t$ is time measured  in the source frame and $t=0$ denotes when the
wind starts. Light rays in this diagram are horizontal lines. 
The jet with $\theta_{\rm j}=10^{\circ}$ and $\Gamma_{\rm j}$=10  is injected at $t = t_{\rm w}$ and is  assumed to continue, with a quasi-steady luminosity $L_{\rm j}$, for a time $t_{\rm j}$. The effects of altering $L_{\rm j}$ for a fixed $t_{\rm j}=1$ s are show in the  {\it top} panel while the consequences of varying $t_{\rm j}$ for a fixed $L_{\rm j} =10^{51}\;{\rm erg/s}$ are depicted in the {\it bottom} panel. Choked jets have $\beta_{\rm j} <\beta_{\rm w}$ while jets with $\beta_{\rm j}\Gamma_{\rm j}<0.5$ are able to break free from the wind but they do so sub-relativistically. In both instances a sGRB will not be triggered.}
\label{fig:stdiag}
\end{figure}

Here we carry out a set of axisymmetric, special relativistic hydrodynamics (SRHD) simulations of the propagation of  a  jet produced in  binary neutron star mergers  and its interaction with the previously spouted  neutrino-driven wind. The simulations employ the \emph{Mezcal} SRHD adaptive mesh refinement code, a parallel, shock capturing code routinely used to simulate the propagation of relativistic jets \citep[e.g.][]{decolle12b, decolle12c}. A detailed description of the code and tests of the SRHD implementation are presented in \citet{decolle12a}. Common to all simulations is the injection of  a steady, spherically symmetric  wind  for a time period $t_{\rm w}$. Immediately after $t_{\rm w}$,  the jet is injected  and the wind power is assumed to decrease with time as $t^{-5/3}$.   For simplicity, we consider the relativistic jet to be uniform with sharp edges and an initial  half-opening angle $\theta_{\rm j}$. 
As a first example, we follow the dynamics of a standard  sGRB  jet expanding through a neutrino-driven wind that was  steadily ejected  over $t_{\rm w}=1$ s before the merger remnant's collapse and the subsequent  formation of the black  hole. The  properties of the neutrino-driven wind have been chosen to roughly  match those envisioned by \citet{dessart09} for a  hot,  differentially rotating merger remnant. Detailed hydrodynamic simulations of the expansion of a relativistic jet  in such a wind  medium are presented in Fig.~\ref{fig:time}, where the  density profiles of the advancing  jet at various times in its evolution are plotted.  As the sGRB jet makes its way through the dense wind, its rate of advance is slowed down and both the shocked jet and shocked wind material advance with a jet head  Lorentz factor $\Gamma_{\rm h} \ll \Gamma_{\rm j}$.   The excess energy must then not collect near the working surface but be accumulated within a cocoon engulfing the jet.  The jet would be expected to escape from the wind region as long as $t_{\rm j}>t_{\rm j,b}$ (equation~\ref{eq:tbl}).  Fig. \ref{fig:time} depicts the  two types of behavior expected  for a steady  sGRB jet expanding in a dense wind with  $\beta_{\rm h} > \beta_{\rm w}$. For  $t_{\rm j}>t_{\rm j,b}$,  the  jet comes out from the edge of the wind at $r_{\rm w}=t_{\rm b} \beta_{\rm w} c$ into an exponentially decreasing atmosphere and into the rarefied ISM beyond it. The jet  will then attain a  bulk Lorentz factor $\Gamma_{\rm j}$ soon after its emergence  and could  lead to a successful sGRB.  For   $t_{\rm j}<t_{\rm j,b}$,  the jet starts to drastically decelerate before it reaches the edge of the wind and  is unable to break free. An accompanying sGRB will not be present in this case.
Some conditions increase the  likelihood  of a successful break-through of the jet.  At a fixed jet luminosity, the risk of stalling is lower when the lifetime of the post-merger remnant is short  or when mass ablation is less efficient. The risk of choking can be also mitigated to a lesser extent by  increasing the jet power.  A similar calculation   to that depicted in  Fig.  \ref{fig:time} can be generated  for jets with varying luminosities  but  similar durations (Fig.~\ref{fig:lum}). For
a relatively low luminosity, the rate of advance of the jet is slowed down drastically and, as a result, the jet is unable to break free (i.e. $t_{\rm j}<t_{\rm j,b}$).  Only when the luminosity is greatly increased would the  jet  be able to successfully emerge  from the wind and subsequently give rise to a sGRB. Note that this is akin to the situation present in long GRBs from collapsars, where the relativistic outflow has to drill through the stellar envelope, with possible observational signatures related to the distribution of angular momentum within the star \citep[e.g.][]{lclrr10}. 

It is evident from the above discussion that the environment of a binary neutron star merger at the time of  black hole formation is complex and that  the properties of the neutrino-driven wind  have a decisive effect on  jet propagation. The  triggering of a sGRB will depends sensitively  on the  jet power and the time over which the jet is injected.  Fig. \ref{fig:stdiag} shows the schematic world-lines of  jets with varying properties propagating through  the same wind environment. The importance of the  longevity of the jet (relative to $t_{\rm w}$) and the significance of the  jet power are clearly illustrated.  

\section{Discussion}\label{dis}

The rate of neutrino-driven mass loss (equation \ref{eq:wind}) emanating from the merger remnant, as argued in  Sections~\ref{ana} and \ref{num}, has significant repercussions  on the appearance of a relativistic jet propagating through it.  But the requirements for  successful break-through are most sensitive  to the duration of the neutrino-driven wind phase ($t_{\rm w}$), which in turn depends on the stability of the resulting hot, differentially rotating post-merger configuration.   If collapse to a black hole and  the ensuing jet production  is significantly delayed (i.e. $t_{\rm w} > 0.1$ s), the majority of jets will not be able to break free from the wind  during the typical duration of a sGRB. On the contrary,  if collapse occurs more promptly (i.e. $t_{\rm w} \ll 0.1$ s), the successful break-through of the jet would take place swiftly enough to allow for  the production of  a  typical sGRB.  

The observed duration distribution  of sGRBs can thus be used to constrain  the longevity of the post-merger remnant, which is currently under  debate. In systems where a stable remnant is formed  \citep[the magnetar model; e.g.][]{metzger2008ex}, jet formation can not be notably delayed  ($\leq 100$ ms) from the onset of the neutrino-driven wind. Otherwise,  a choked jet would result without exception. This is because  in stable (hyper-massive) neutron stars, the neutrino-driven wind is expected to continue for at least a diffusion timescale, which is commensurate  with the  duration of the longest lasting sGRB. Effective sGRB production under such circumstances would not only be afflicted by baryon contamination (equation~\ref{eq:bpop}) and wind confinement  (equation~\ref{eq:lumreq}) but also by the time delay between the onset of neutrino-driven mass ablation and jet formation. Whatever one's view of the relative merits of the magnetar  and the black hole plus debris disk models in producing  ultra-relativistic outflows, it is clear that the presence of a long lived, dense wind, a common feature in the magnetar model,  severely hinders the successful production of a sGRB.  

The shocks responsible for producing the $\gamma$-rays must surface after the  jet has broken free from the neutrino-driven wind. For a large subset of compact merger progenitors, with the exception of black hole neutron star mergers, a hyper-massive remnant will be formed. A neutrino-driven wind will thus persist to obstruct the progress of the jet. 
The cocoon would be able to  collimate the jet  provided that $\tilde{L} < \theta_{\rm j}^{-4/3}$ \citep{bromberg11}. In the simplest case considered here of a wind whose properties do not vary over its lifetime,  the jet's head velocity is constant and the cocoon is unable to compensate  for the jet's expansion (Figures~\ref{fig:time} and~\ref{fig:lum}). Collimation is  seen to increase with decreasing $k$ for $k \leq 2$ where $\rho_{\rm w} \propto r^{-k}$. While writing this paper we became aware of a recent preprint \citep{nagakura2014}, in which the interaction of a jet with the previously dynamically ejected material (which is independent of the neutrino-driven wind we consider here) in a  binary merger \citep{hotokezaka13b}  is used to  argue as an effective mechanism for the collimation of a sGRB jet. 

The energy supplied by the jet exceeds that imparted to the swept-up wind  material by a factor  $\beta_{\rm h}/\beta_{\rm j} < 1$. The excess  energy must then not gather near the working surface but be deposited within a cocoon surrounding the jet (Figures~\ref{fig:time} and \ref{fig:lum}). Unless there is violent mixing of baryons from the wind,  the build-up of energy to baryon-rest-mass in the cocoon will be given approximately by $\approx \Gamma_{\rm h}$. As soon as the jet reaches the edge of the wind, $r_{\rm w}$, the cocoon material would itself be able to break-out and expand through the wind along the direction of least resistance, which is likely to be along the jet's  axis.  Beyond $r_{\rm w}$, the external pressure drops steeply, and the cocoon material will expand freely with $\Gamma_{\rm c}  \propto r/r_{\rm w}$ \citep{ramirez-ruiz2002}. If its unhampered transverse  expansion 
starts just outside $r_{\rm w}$, where the Lorentz factor of the cocoon material  is only a few, then it will spread over a wide angle. 

In the case of a successful break-through of the relativistic jet, the outflowing cocoon could result in potentially interesting and observable phenomenon \citep{ramirez-ruiz2002, thompson2007, morsony2007}. The relativistic material that accumulated in the cocoon could have an energy comparable to that of the jet when $t_{\rm j} \gtrsim t_{\rm j,b}$. Not only would a typical  sGRB be detectable, followed by a standard afterglow, but also there could be additional emission  before and after the main event when the cocoon material becomes transparent and  when it decelerates.  While the main afterglow radiation will be produced by the slowing down of the jet as in the usual case, prompt X-ray  emission at early stages  could  be caused by the deceleration of the cocoon blast wave, which is expected  to be less energetic. This could resemble the so called {\it extended emission} in sGRBs \citep{norris2006}, in particular  if it is slowly varying.   There could also be precursor signatures \citep{ramirez-ruiz2001,troja2010} which are not associated with internal dissipation in the jet, but with the dynamics of the cocoon fireball.  The $\gamma$-ray signal  emerging from the cocoon fireball as it becomes transparent  will most likely appear as a transient signal before  the beginning of the main burst \citep{ramirez-ruiz2002, suzuki2013}, where the observed variability time scale would be  related to the typical size of the shocked plasma region containing the photon field: $\Delta \approx r_{\rm w}$ for $r/\Gamma_{\rm c}^2 <\Delta$. The detection of these prompt  signatures would be a test of the  neutron star binary merger  model and the precise measurement of the time delay between emissions may help constrain the duration and properties of the neutrino driven wind phase.

If we were to venture a general classification scheme for GRBs, on the hypothesis that the central engine involves a black hole formed in double neutron star mergers, we would obviously expect the disk and black hole mass \citep{lee2005b,oechslin2006,giacomazzoetal13}, the angular momentum of the black hole and the orientation relative to our line of sight to be essential parameters.  It is then necessary, within such a model,   to identify sGRBs  with  merging  systems  for which   black hole  formation occurs promptly ($\leq 100$ ms) as any moderate delay  at the hyper-massive neutron star stage would result in a choked  jet.

\acknowledgements
We thank C. Holcomb, L. Rezzolla, T. Janka,  B. Giacomazzo,  L. Lehner, B. Metzger, C. Palenzuela, L. Roberts, S. Rosswog  for insightful discussions and  acknowledge financial support from the Packard Foundation, NSF (AST0847563),  UCMEXUS (CN-12-578), CONACyT (101958) and PAPIIT-UNAM (IG100414 and IA101413-2).

\end{document}